\documentclass[12pt]{article} 
\usepackage{a4} 
\usepackage{epsfig}
\usepackage{amssymb} 
\usepackage{amsmath} 
\usepackage{amsfonts}

\def\vectau{{\pmb{\tau}}}

\newtheorem {theorem}{Theorem}[section]
\newtheorem {lemma}[theorem]{Lemma}

\begin{document}
\title{
  \vskip 15pt
  {\bf \Large \bf Finkelstein-Rubinstein constraints for the Skyrme
  model with pion masses}
  \vskip 10pt}
\author{
Steffen Krusch\thanks{S.Krusch@kent.ac.uk } \\[5pt]
{\normalsize {\sl Institute of Mathematics and Statistics}}\\
{\normalsize {\sl University of Kent,
Canterbury CT2 7NF}}\\
{\normalsize {\sl United Kingdom}}
}

\date{September 2005}

\maketitle

\begin{abstract}
The Skyrme model is a classical field theory modelling the strong
interaction between atomic nuclei. It has to be quantized in order to
compare it to nuclear physics. When the Skyrme model is
semi-classically quantized it is important to take the
Finkelstein-Rubinstein constraints into account. Recently, a simple
formula has been derived to calculate the these constraints for
Skyrmions which are well-approximated by rational maps. However, if a
pion mass term is included in the model, Skyrmions of sufficiently
large baryon number are no longer well-approximated by the rational
map ansatz. This paper addresses the question how to calculate
Finkelstein-Rubinstein constraints for Skyrme configurations which are
only known numerically.  
\end{abstract}

\vspace{2cm} 
\centerline{PACS-number: 12.39.Dc}
\vspace{2cm}
\begin{tt}
\pageref{lastref} pages, 1 figure
\end{tt}

\newpage
\setcounter{page}{1} 

\section{Introduction}

The Skyrme model is a classical model of the strong interaction
between atomic nuclei \cite{Skyrme:1961vq}.
In order to compare the
Skyrme model with nuclear physics we have to understand the classical
solutions and then quantize the model. The classical solutions have a
surprisingly rich structure. Configurations in the Skyrme model are
labelled by a topological winding number which can be interpreted as the
baryon number $B$. Static minimal energy configurations for a given
$B$ are known as Skyrmions. The $B=1$ Skyrmion has spherical symmetry,
the $B=2$ Skyrmion has axial symmetry and for $B>2$ Skyrmions have
various discrete symmetries, see \cite{Battye:2001qn} and references
therein. The Skyrme model depends on a parameter which corresponds to
the pion mass $m_\pi$. For $m_\pi = 0$ all the Skyrmions for
$B \le 22$ were found to be shell-like configurations with discrete
symmetries \cite{Battye:2001qn}. Such configurations are very well
described by the rational map ansatz \cite{Houghton:1998kg}.
However, if the value of the pion mass is increased to its
physical value or higher, then for high enough baryon number
shell-like solutions are no longer the minimal energy solutions
\cite{Battye:2004rw}. 

In \cite{Adkins:1983ya,Adkins:1984hy} Adkins et al. quantized the
translational and rotational zero modes of the $B=1$ Skyrmion for zero
and nonzero pion mass respectively and obtained good agreement with
experiment. A subtle point is that Skyrmions can be quantized as
fermions as has been shown in \cite{Finkelstein:1968hy}. Solitons in
scalar field theories can consistently be quantized as fermions
provided that the fundamental group of configuration space has a
${\mathbb Z}_2$ subgroup generated by a loop in which two identical
solitons are exchanged. All loops in configuration space give rise to
so-called Finkelstein-Rubinstein constraints which depend on whether a
loop in configuration space is contractible or not. In particular,
symmetries of classical configurations induce loops in configurations
space. After quantization, these loops give rise to constraints on the
wave function. 

The $B=2$ Skyrmion with axial symmetry was quantized in
\cite{Braaten:1988cc, Verbaarschot:1987au} using
the zero mode quantization. Later, the approximation was improved by
taking massive modes into account \cite{Leese:1995hb}. 
The $B=3$ Skyrmion was first quantized in \cite{Carson:1991yv} and the
$B=4$ Skyrmion \cite{Walhout:1992gr}. Irwin 
performed a zero mode quantization for $B=4-9$ \cite{Irwin:1998bs}
using the monopole moduli space as an approximation for the Skyrmion
moduli space. 
The physical predictions of the Skyrme model for various baryon numbers 
were also discussed in \cite{Kopeliovich:2001yg}.
Recently, the Finkelstein-Rubinstein constraints have 
been calculated for Skyrmions which are 
well-approximated by the rational map ansatz \cite{Krusch:2002by}, and
which we shall call rational map Skyrmions. In
this case, the Finkelstein-Rubinstein constraints are given by a
simple formula. This formula is also valid if the Skyrme configuration can be
deformed into a rational map Skyrmion while preserving the
relevant symmetries. However, this is not always possible. 
The aim of this paper is to show how to calculate
Finkelstein-Rubinstein constraints for more general configurations. 

This paper is organized as follows. In section \ref{rationalmaps} we
first describe the rational map ansatz. Then we discuss the
Finkelstein-Rubinstein constraints. Finally, we derive some
constraints on the symmetries which are compatible with a rational map
Skyrmion. 
In section \ref{truncated} we first introduce a truncated rational map
ansatz which describes well-separated rational map Skyrmions. Then we
calculate the Finkelstein-Rubinstein constraints for this class of
Skyrme configurations. In the following section, we describe how to
calculate the Finkelstein-Rubinstein constraints for a minimal energy
configuration which is only known numerically. We also give an
example. In section \ref{evenB} we derive constraints on possible
symmetries in order to make predictions about ground states for
Skyrmions with even baryon number. We end with a conclusion.

\section{Skyrmions and Rational Maps}
\label{rationalmaps}

In this section, we first recall the some basic facts about the Skyrme
model. Then we describe the rational map ansatz. We then
discuss how to quantize a Skyrmion as a fermion. Finally, we derive
which symmetries are compatible with a rational map Skyrmion of a given baryon
number.

\subsection{The rational map ansatz}

The Skyrme model is a classical field theory of pions. The basic field
is the $SU(2)$ valued field $U({\bf x},t)$ where ${\bf x} \in {\mathbb
  R}^3$. The static solutions can be obtained by varying the
following energy
\begin{equation}
\label{energy}
E = \int \left( - \frac{1}{2} {\rm Tr} (R_i R_i) 
- \frac{1}{16} {\rm Tr} ([R_i,R_j][R_i,R_j]) 
- m^2 {\rm Tr} (U-1) \right) d^3 x,
\end{equation}
where $R_i = (\partial_i U) U^\dagger$ is a right invariant $su(2)$
valued current, and $m$ is a parameter proportional to the pion mass
$m_\pi$, 
\cite{Battye:2004rw}. In order to have finite energy, Skyrme fields
have to take a constant value, $U(|{\bf x}| = \infty) = 1$, at
infinity, and such maps are characterized by an integer-valued winding
number. This topological charge is interpreted as the baryon number
and is given by the following integral
\begin{equation}
\label{B}
B= - \frac{1}{24 \pi^2} \int \epsilon_{ijk} {\rm Tr}(R_i R_j R_k) d^3 x.
\end{equation} 
We will denote the configuration space of Skyrmions by $Q$. $Q$ splits
into connected components $Q_B$ labelled by the topological
charge. Furthermore, the energy of configurations in $Q_B$ is bounded
below by $E \ge 12 \pi^2 B$ \cite{Faddeev:1976pg}.

The minimal energy solutions have been calculated in the massless
case, $m=0$, for all $B \le 22$
\cite{Battye:2001qn}. The solutions are shell-like structures which
are very well approximated by the rational map ansatz
\cite{Houghton:1998kg} which we will now describe. 

The main idea is to write Skyrme fields which can be thought of as
maps from $S^3 \to S^3$ 
in terms of rational maps which are holomorphic maps from $S^2
\to S^2$. In algebraic topology, such a construction is known as a
suspension. First, we introduce polar coordinates $(r,\theta,\phi)$
and note that the angular coordinates can be related to the complex
plane $z$ by the stereographic projection $z = {\rm e}^{i \phi} \tan
\frac{\theta}{2}$. Then the Skyrme field can be written as
\begin{equation}
\label{ansatz}
U(r,z) = \exp\left( \frac{i f(r)}{1 + |R|^2} 
\left(
\begin{array}{cc}
1 - |R|^2 & 2 {\bar R} \\
2 R & |R|^2 - 1 
\end{array}
\right)
\right),
\end{equation}
where the profile function $f(r)$ is a real function satisfying the
boundary conditions $f(0) = \pi$ and $f(\infty) = 0$. The map $R = R(z)$ is
the eponymous rational map. It can be written as the quotient of two
polynomials $p(z)$ and $q(z)$ which satisfy
${\rm max}(\deg(p(z)),\deg(q(z))) = B$ 
and $p(z)$ and $q(z)$ have no common factors. Here
$\deg$ denotes the polynomial degree. 
The ansatz (\ref{ansatz}) can be inserted into the energy
(\ref{energy}) and we obtain
\begin{equation}
\label{ratenergy}
E = 4 \pi \int \left( r^2 {f^\prime}^2 + 
2 B \left({f^\prime}^2 + 1\right) \sin^2 f + 
{\cal I} \frac{\sin^4 f}{r^2} + 2 m^2 r^2 \left(1 - \cos f \right)
\right) dr,
\end{equation}
where
\begin{equation}
{\cal I} = \frac{1}{4 \pi} \int \left( 
\frac{1 + |z|^2}{1 + |R|^2} \left| \frac{dR}{dz} \right| \right)^4
\frac{2i~ dz~d{\bar z}}{(1+|z|^2)^2}.
\end{equation}
To minimize the energy (\ref{ratenergy}) one first determines the
rational map which minimizes ${\cal I}$ and then calculates the shape
function $f(r)$ numerically by solving the corresponding
Euler-Lagrange equation. The rational maps which minimize ${\cal I}$
have been determined numerically in \cite{Battye:2001qn,
  Battye:2002wc} for all $B \le 40$. 
Note that the restriction that $R(z)$ is a
holomorphic map can be lifted and a generalized rational map ansatz
can be introduced \cite{Houghton:2001fe}. This generalized ansatz has
been shown to improve the energy significantly for $B \le 4$, and it
also captures the singularity structure of Skyrmions better. 
However, it is difficult to use for
higher baryon number, and from the point of view of discussing
symmetries the original rational map ansatz is sufficient.

The rational map ansatz gives a good approximation to the energy of
a Skyrmion and also gives a very accurate prediction of its
symmetry \cite{Battye:2001qn}. 
By symmetry we mean that a rotation in space followed by a
rotation in target space leaves the Skyrmion invariant. Namely,
\begin{equation}
U({\bf x} ) = A U \left(D(A^\prime) {\bf x} \right) A^\dagger,
\end{equation}
where $A$ and $A^\prime$ are $SU(2)$ matrices and $D(A^\prime)$ is the
associated $SO(3)$ rotation. It is
therefore important to understand how a rational map transforms under
rotations in space and target space. It can be shown that
\begin{equation}
R(z) \mapsto {\tilde R}(z) = M_A (R(M_{A^\prime}(z))),
\end{equation}
where $M_A$ and $M_{A^\prime}$ are M\"obius transformations. See
\cite{Krusch:2002by} for further details.

\subsection{Finkelstein-Rubinstein constraints}

In the following, we recall the ideas of Finkelstein and Rubinstein
\cite{Finkelstein:1968hy} on how to quantized a scalar field theory and obtain
fermions. For further details see \cite{Krusch:2002by, Krusch:2005bn}. 
The main idea is to define a wave function on the covering space of
configuration space. Recall that the configuration space $Q$ of the
Skyrme model splits into connected components labelled by the degree
$B$, and will be denoted by $Q_B$.
The fundamental group of each component of the configuration space $Q$ 
is $\pi_1(Q_B) = \mathbb{Z}_2$. Therefore, the covering space ${\tilde
  Q}_B$ of each component is a double cover. In order to have
fermionic quantisation we have to impose the condition that if two
different points $p_1, p_2 \in {\tilde Q}_B$ correspond to the same
point $p \in Q_B$ then the wave function $\psi: {\tilde Q}_B \to
{\mathbb C}$ has to satisfy
\begin{equation}
\label{fermionic}
\psi(p_1) = - \psi(p_2).
\end{equation}
The points $p_1, p_2 \in {\tilde Q}_B$ can be interpreted as two paths in
configuration space. The condition that $p_1 \neq p_2$ implies that
$p_1$ and $p_2$ differ by a noncontractible loop. Every symmetry of a
classical configuration gives rise to a loop in configuration space.  
In particular, we are interested in symmetries given by 
a rotation by $\alpha$ in space
followed by a rotation by $\beta$ in target space. 
This leads to the
following constraint on the wave function $\psi$: 
\begin{equation}
\label{condition}
\exp\left( i \alpha {\bf n} \cdot {\hat {\bf J}} \right)
\exp\left( i \beta {\bf N} \cdot {\hat {\bf I}} \right)
\psi = \chi_{FR} \psi,
\end{equation}
where ${\bf n}$ is the direction of the rotation axis in space, ${\bf
N}$ is the rotation axis in target space, a ${\hat {\bf J}}$ and
${\hat {\bf I}}$ are the angular momentum operators in space and
target space, respectively.\footnote{For a more detailed discussion on
  body fixed and space fixed angular momentum operators in this context see
  \cite{Krusch:2005bn}.} 
Note that rotations in target space will also be called isorotations. 
The Finkelstein-Rubinstein phase
$\chi_{FR}$ enforces the condition (\ref{fermionic}) and satisfies
\begin{equation}
\chi_{FR} =
\left\{
\begin{array}{cl}
1 &~~~{\rm if~the~induced~loop~is~contractible}, \\
& \\
-1 &~~~{\rm otherwise}.
\end{array}
\right.
\end{equation}
Here is a good place to summarize some important and well-known results. 
Giulini showed that a $2 \pi$ rotation of a Skyrmion gives rise to
$\chi_{FR} = (-1)$ 
if and only if the baryon number $B$ is odd, \cite{Giulini:1993gd}.
Finkelstein and Rubinstein showed in \cite{Finkelstein:1968hy} that a
$2 \pi$ rotation of a Skyrmion of degree $B$ is homotopic to an
exchange of two Skyrmions of degree $B$. This also implies that an
exchange of two identical Skyrmions gives rise to $\chi_{FR} = (-1)$
if and only if their baryon number $B$ is odd. 
In \cite{Krusch:2002by} it was shown that a $2 \pi$ isorotation of a 
Skyrmion also gives rise to $\chi_{FR} = (-1)$ if and only if the 
baryon number $B$ is odd.
These results agree
with the physical intuition since atomic nuclei can be modelled by
interacting point-like fermionic particles. 

\subsection{Symmetries of rational maps}

Shell-like Skyrmions are described very well using the rational map 
ansatz. If a rational map Skyrmion of degree $B$ is symmetric under a
rotation by $\alpha$ followed by an isorotation by $\beta$ then the 
Finkelstein-Rubinstein phase of this symmetry is given by 
\begin{equation}
\begin{array}{ccc}
\label{N}
\chi_{FR} = (-1)^N & {\rm where} & 
N =  B \left( B \alpha - \beta \right)/(2 \pi),
\end{array}
\end{equation}
which has been proven in \cite{Krusch:2002by}. For $B>2$ all the known
Skyrmions are invariant under discrete subgroups,  
so they contain cyclic groups as subgroups. Let $C_n^k$ be a cyclic group 
of order $n$ which is generated by a rotation by $\alpha = 2 \pi /n$ 
followed by an isorotation by $\beta = 2 \pi k/n$ where $-n < k \le n$.
Equation (\ref{N}) imposes a constraint on the values of $B$ which are 
compatible with a given $C_n^k$ symmetry. Namely, $N$ has to be an 
integer. 
A stronger constraint can be derived if we work directly with rational 
maps.

\begin{lemma}
A rational map of degree $B$ can have a $C_n^k$ symmetry if and only if 
$B \equiv 0 \mod n$ or $B \equiv k \mod n$.
\end{lemma}

\noindent {\bf Proof:} \\
Without loss of generality consider rational maps with boundary
condition $R(\infty) = \infty$ and assume that the $C_n^k$ symmetry 
corresponds to a rotation around the third axis in space
followed by a rotation around the negative third axis in target
space. With this choice of axes, the boundary conditions are preserved 
by the relevant rotation and also by the relevant isorotation, 
and the sign choice corresponds to the sign choice
for (\ref{N}) in \cite{Krusch:2002by}.  
The rational map $R(z) = p(z)/q(z)$ can be written as
\begin{equation}
\label{Rat}
R(z) = \frac{z^B + a_{B-1} z^{B-1} + \dots + a_0}
{b_{B-1}z^{B-1}+\dots+b_0}
\end{equation}
where the polynomials $p(z)$ and $q(z)$ have no common factors.
The $C_n^k$ symmetry condition is given by
\begin{equation}
R(z) = {\rm e}^{-2 \pi k i/n} R({\rm e}^{2 \pi i/n} z).
\end{equation} 
Note that a $C_n^{n-k}$ rotation can be interpreted as a $C_n^{-k}$
rotation followed by a $2 \pi$ isorotation. Since for a $2 \pi$
isorotation the Finkelstein-Rubinstein phase is simply given by 
$\chi_{FR} = (-1)^B$, we can restrict our attention to $k = 0, \dots, n-1$.

First we show existence. Let $B \equiv k \mod n$, so $B = nl +
k$. Then the rational  map
\begin{equation}
\label{rat1}
R(z) = \frac{z^k r(z^n)}
{s(z^n)}
\end{equation}
where $r(z)$ is a polynomial of degree $l$ and $s(z)$ is a polynomial
of at most degree $l$. For $k=0$ the degree of $s(z)$ has to be less
than $l$ in order to respect the boundary conditions $R(\infty)=\infty$.
This rational map is invariant under $C_n^k$. To make sure that it is
a rational map of degree $B = nl+k$ the polynomials $r(z^n)$ and
$s(z^n)$ are required not to have any common factors. 
Furthermore, for $k \neq 0$, we need to impose
$s(0) \ne 0$, since the polynomial $p(z)$ has a zero at $z=0$. 
For $k=0$, we also impose the condition $s(0) \ne 0$, and we will
discuss the case $s(0) = 0$ in the next paragraph.
The simplest example of such a rational map is
\begin{equation}
R(z) = z^B.
\end{equation}
Hence a rational map of degree $B$ with $C_n^k$ symmetry exists for $B 
\equiv k \mod n$.

Similarly, let $B \equiv 0 \mod n$, so $B = nl$. Again, we
only consider $k=0, \dots, n-1$. Then the rational map  
\begin{equation}
\label{rat2}
R(z) = \frac{r(z^n)}
{z^{n-k} s(z^n)},
\end{equation}
is invariant under $C_n^k$. Here $r(z)$ is again a polynomial of
degree $l$ and $s(z)$ is a polynomial of at most degree
$l-1$ which has no common factors with $r(z)$. 
Furthermore, we also require $r(0) \ne 0$. 
One example of such a rational map is
\begin{equation}
R(z) = \frac{z^B+1}{z^{n-k}}.
\end{equation} 
Hence a rational map of degree $B$ with $C_n^k$ symmetry exists for $B 
\equiv 0 \mod n$. This completes the proof of existence. The
classification of rational maps into type (\ref{rat1}) and type
(\ref{rat2}) will become useful in section \ref{evenB}.

Now, we assume that the rational map (\ref{Rat}) is invariant under $C_n^k$. 
In homogeneous coordinates, the rational map  $R(z)$ is given by 
$[p(z),q(z)] \in {\mathbb C}P^1$ 
subject to the relation that $[p(z),q(z)] = [\lambda p(z), 
\lambda q(z)]$ for any complex number $\lambda 
\neq 0$. Under the symmetry $C_n^k$, the polynomials $p(z)$ and $q(z)$ 
in equation (\ref{Rat}) transform as 
\begin{equation}
\begin{array}{ccc}
p(z) &\mapsto& \lambda {\rm e}^{-2 \pi i k/n} p({\rm e}^{2\pi i/n} z), \\
q(z) &\mapsto& \lambda q({\rm e}^{2 \pi i/n} z), 
\end{array}
\end{equation}
for $\lambda \neq 0$. Assuming the rational map is invariant under 
$C_n^k$ leads to the following constraints on the coefficients.
\begin{equation}
\label{constraint}
\begin{array}{ccc}
a_m &=& \lambda {\rm e}^{2 \pi i (m-k)/n} a_m, \\  
b_m &=& \lambda {\rm e}^{2 \pi i m/n} b_m.
\end{array}
\end{equation}
In order to have a rational map of degree $B$, $p(z)$ and $q(z)$ cannot 
have any common factors. This implies that $a_0$ and $b_0$ cannot both be 
zero. First, assume $a_0 \neq 0$. Then equation (\ref{constraint}) implies 
that
\begin{equation}
\lambda = {\rm e}^{2 \pi i k/n}.
\end{equation} 
The coefficient of the highest power in the numerator is also not allowed 
to vanish which implies
\begin{equation}
{\rm e}^{2 \pi i B/n} = 1,
\end{equation}
so that $B \equiv 0 \mod n$.
Now, consider that $b_0 \neq 0$ which implies 
\begin{equation}
\lambda = 1.
\end{equation}
Again, the coefficient of the highest power in the numerator is not 
allowed to vanish. Therefore, 
\begin{equation}
{\rm e}^{2 \pi i (B-k)/n} = 1,
\end{equation}
so that $B \equiv k \mod n$, which completes the proof.  \hfill $\square$

The lemma is more restrictive than the condition that $N$ is an integer. 
For example $B(B-k) \equiv 0 \mod n$ suggests that a $C_4^0$ symmetry is 
possible for $B=2$. However, since $B \equiv 2 \mod 4$ our lemma excludes 
such a symmetry.

\section{A truncated rational map ansatz}
\label{truncated}

In \cite{Krusch:2002by} the Finkelstein-Rubinstein constraints were
calculated for Skyrmions which are well approximated by the rational
map ansatz. In this section, we calculate the Finkelstein-Rubinstein
constraints for Skyrme configurations $U({\bf x})$ which are given by
a truncated rational map ansatz defined as follows. Let 
$U_{B_i}({\bf x})$ be a Skyrme configuration of degree $B_i$ which is
given by (\ref{ansatz}) and the shape function $f(r)$ is a smooth,
decreasing function which satisfies $f(0) = \pi$ and $f(r) = 0$ for $r
\ge L$. Then the Skyrme configuration is given by
\begin{equation}
U({\bf x}) = 
\left\{
\begin{array}{lll}
U_{B_i}({\bf x} - {\bf X}_i) & ~~~~~~ & {\rm for}~|{\bf x} - {\bf
  X}_i| < L, \\
1 & & {\rm otherwise}.
\end{array}
\right.
\end{equation}
From formula (\ref{B}) it is obvious that the configuration $U({\bf x})$ 
has the degree $B = \sum\limits_i B_i$.
The parameters ${\bf X}_i$ are the positions of the Skyrmions, and we
assume that $|{\bf X}_i - {\bf X}_j| > 2L$ for $i \ne j$. Such an
ansatz provides reasonable initial conditions for numerical
simulations \cite{PC1}. 
A related ansatz is the product ansatz. This ansatz produces Skyrme
configurations which are closer in energy to the true
solutions. Topologically, these two ans\"atze are equivalent. However,
the product ansatz has the disadvantage that it is non-commutative,
since in general $U_1 U_2 \ne U_2 U_1$ for $SU(2)$ matrices $U_1$ and
$U_2$, so that it is slightly more difficult to discuss
symmetries. Therefore, we restrict our attention to the truncated
rational map ansatz.

Consider a configuration $U({\bf x})$ which is invariant under
$C_n^k$. The symmetry relates different Skyrmions $U_{B_i}$ with each
other. 
For each individual Skyrmion $U_{B_i}$ there are two possibilities.
Either the centre of this Skyrmion lies on
the symmetry axis or it is one constituent of a regular $n$-gon of 
Skyrmions which transform into each other under the symmetry. 

\subsection{Skyrmions centred on the symmetry axis}
\label{axial}

Assume two Skyrmions with baryon number $B_1$ and $B_2$ have a common 
$C_n^k$ axis of symmetry, say the $x_3$ axis, and are centred around
the origin and the point $P = (0,0,c)$ for $c>L$. 
Then the symmetry loop is homotopic to a 
product of two loops, each acting only on one Skyrmion. This can be
seen as follows. The configuration can be written as
\begin{equation}
U({\bf x}) =
\left\{ 
\begin{array}{ccl}
U_1({\bf x}) &~~~~&{\rm for~} |{\bf x}| < L,\\
U_2({\bf x}) && {\rm for~} |{\bf x} - (0,0,c)| < L,\\
1&& {\rm otherwise.}
\end{array}
\right.
\end{equation}
Under a rotation, the configuration transforms as
\begin{equation}
A(\beta) U(D(\alpha){\bf x}) A^\dagger(\beta) =
\left\{ 
\begin{array}{ccl}
A(\beta)U_1(D(\alpha){\bf x})A^\dagger(\beta) &~~~~&{\rm for~} |{\bf x}| < L,\\
A(\beta)U_2(D(\alpha){\bf x})A^\dagger(\beta) 
&& {\rm for~} |{\bf x} - (0,0,c)| < L,\\
1&& {\rm otherwise}.
\end{array}
\right. 
\end{equation} 
Here $A(\beta)$ is a rotation in target space acting by conjugation
and $D(\alpha)$ is a rotation around the $x_3$ axis.
Note in particular, that the vacuum is invariant under isorotation. 
The map $H: [0,1] \times [0,1] \to Q_B: (s,t) \mapsto H(s,t)$ 
provides a homotopy such that $H(0,t)$ is a $C_n^k$ 
rotation of the whole configuration while $H(1,t)$ corresponds to a
loop which first rotates one Skyrmion and then the other.
\begin{equation}
\label{homotopy}
H(s,t) = 
\left\{
\begin{array}{cl}
A(h_1(s,t) \beta) U_1(D(h_1(s,t)\alpha){\bf x})A^\dagger(h_1(s,t) \beta) 
&{\rm for~} |{\bf x}| < L,\\
A(h_2(s,t) \beta) U_2(D(h_2(s,t)\alpha){\bf x})A^\dagger(h_2(s,t) \beta) 
& {\rm for~} |{\bf x} - (0,0,c)| < L,\\
1& {\rm otherwise},
\end{array}
\right.
\end{equation}
where 
\begin{equation}
h_1(s,t) = 
\left\{
\begin{array}{ccl}
t/(1-s/2) & ~~~~~& {\rm for~} 0\le t < 1-s/2, \\
1 & & {\rm for~} s \le t \le 1,
\end{array}
\right.
\end{equation}
and similarly
\begin{equation}
h_2(s,t) = 
\left\{
\begin{array}{ccl}
0  & ~~~~~& {\rm for~} 0\le t < s/2, \\
(t-s/2)/(1-s/2) & & {\rm for~} s/2 \le t \le 1.
\end{array}
\right.
\end{equation}
Therefore, the Finkelstein-Rubinstein phase can be calculated as a product
of the two loops of the individual Skyrmions. 
Since we are assuming that the $B_1$ and 
the $B_2$ Skyrmion are both well-described by the rational map ansatz
we can apply formula (\ref{N}) and obtain $\chi_{FR} = (-1)^N$ where
\begin{eqnarray}
\nonumber
N &=& B_1(B_1 \alpha - \beta) /(2 \pi) + B_2 (B_2 \alpha - \beta) /(2 \pi), 
\\
\label{Naxial}
&=& (B_1^2+B_2^2) \alpha/(2 \pi) - (B_1 + B_2) \beta /(2 \pi).
\end{eqnarray}
Naive application of formula (\ref{N}) gives an incorrect 
result, namely $\chi_{FR} = (-1)^{\tilde N}$ where
\begin{equation}
{\tilde N} = (B_1+B_2)^2 \alpha/(2 \pi) - (B_1 + B_2) \beta /(2 \pi).
\end{equation}
This ${\tilde N}$ is no longer well defined. Let $B_1=1$ and $B_2=1$ be 
invariant under a $C_3^1$ rotation. Then $N = 0$, but ${\tilde N} = 2/3$.
Also, consider $B_1 = 1$ and $B_2 = 1$ with symmetry $C_2^1$ then 
$N = 0$, but ${\tilde N} = 1$.

In this context, it is worth mentioning another interesting ansatz for
Skyrmions, namely, the multi-shell ansatz by Manton and Piette
\cite{Manton:2000kj}. The main idea is to construct multiple
concentric shells of Skyrmions, where each shell is given by the usual
rational map ansatz. The multi-shell ansatz can then be written as
\begin{equation}
U({\bf x}) = \left\{
\begin{array}{ccc}
\exp \left(i f(r) {\bf n}_{R_1} \cdot {\vectau} \right)
& ~~~~~~~~ & {\rm for~} 2\pi \ge f(r) > \pi, \\
&& \\
\exp \left(i f(r) {\bf n}_{R_2} \cdot {\vectau} \right)
& & {\rm for~} \pi \ge f(r) > 0,
\end{array}
\right. 
\end{equation}  
where $f(r)$ is a monotonically decreasing function with boundary
conditions $f(0) = 2 \pi$ and $f(\infty) = 0$. Here $R_1$ and $R_2$
are two rational maps of degree $B_1$ and $B_2$, respectively. 
The degree of such a Skyrme configuration is $B = B_1 + B_2$. 
Note that for $f(r_0) = \pi$ the Skyrme field takes the value 
$U(r_0) = -1$ which is
invariant under rotations and isorotations. As before, we can 
split a symmetry loop into two loops, each of which only acting on the
inner or the outer Skyrmion, similar to (\ref{homotopy}).
Therefore, formula (\ref{Naxial}) is also valid in this case.

\subsection{Configurations of $n$ Skyrmions related by symmetry}

\begin{lemma} 
\label{nN}
For $n$ Skyrmions of degree $B$ which are related by a $C_n^k$
symmetry the Finkelstein-Rubinstein phase is $\chi_{FR} =(-1)^N$ where
\begin{equation}
N = B \left(n B - k \right).
\end{equation}
\end{lemma}

\noindent {\bf Proof:}

The relevant Skyrme configuration corresponds to a regular $n$-gon of 
Skyrmions.
The loop $L$ which is induced by the $C_n^k$ symmetry has two effects on
this configuration. 
Each Skyrmion is rotated and isorotated by
$C_n^k$. Furthermore, the Skyrmions are exchanged via the permutation
$(1 2 \dots n)$, keeping the orientation in space and target space fixed. 
Therefore, the loop $L$ can be divided into $n$ rotation and isorotation 
loops for
each Skyrmion and a permutation loop, which in turn can be split up
into $n-1$ exchanges of two Skyrmions. The individual Skyrmions are
given by rational maps so we can apply formula (\ref{N}) for each
rotation and isorotation. 
Furthermore, Finkelstein and Rubinstein have shown that an
exchange of two Skyrmions of degree $B$ is homotopic to a $2 \pi$
rotation of a Skyrmion of degree $B$ and we can again apply formula
(\ref{N}). Note that $N$ in formula (\ref{N}) is only well-defined
modulo $2$.
The Finkelstein-Rubinstein phase for the loop $L$ is then
given by 
\begin{equation}
\chi_{FR} = \left(\chi_{FR}({\rm rotation/isorotation})\right)^n 
\left(\chi_{FR}({\rm single~exchange})\right)^{n-1}.
\end{equation}
Using (\ref{N}) this gives rise to 
\begin{eqnarray}
N &=& n B\left(B/n -k/n\right) + (n-1) B^2, \\
  &=& n B^2 - B k, 
\end{eqnarray}
which completes the proof. \hfill $\square$

A more heuristic way of understanding the result of lemma \ref{nN} is
the following. Consider a regular $n$-gon of Skyrmions of degree $B$ which
transform into each other under a $C_n^k$ symmetry.
Intuitively, this configuration can be deformed into a torus of degree $n B$ 
under a homotopy which preserves the $C_n^k$ symmetry. Then the
Finkelstein-Rubinstein phase can be calculated with formula (\ref{N})
and we obtain 
\begin{equation}
\chi_{FR} = (-1)^{B \left(nB - k \right)},
\end{equation} 
as above.

\subsection{General configurations}

Given a general configuration in the truncated rational map ansatz,
which is symmetric under $C_n^k$, we split up the configuration into
regular $n$-gons of Skyrmions which transform into each other and Skyrmions
which are on the symmetry axes. Assume that there are $l$ regular 
$n$-gons of Skyrmions with degree $B_i$ for $i=1,\dots, l$ and $m$ Skyrmions
of degree ${\tilde B}_i$ for $i=1,\dots, m$ which are located on the
symmetry axis. Then the Finkelstein-Rubinstein phase for this symmetry
is given by 
\begin{equation}
\label{generalN}
\chi_{FR} = (-1)^N  ~~{\rm where}~~ 
N = \sum\limits_{i=1}^{l} B_i \left(n B_i - k \right) 
+ \sum_{i=1}^{m} {\tilde B}_i\left({\tilde B}_i - k \right)/n.
\end{equation}
This formula follows by constructing a homotopy between the $C_n^k$ 
symmetry loop and a product of loops for the individual groups of
Skyrmions as in section \ref{axial}.

\section{How to calculate Finkelstein-Rubinstein constraints from 
numerical configurations}
\label{numeric}

In this section, we describe how to calculate the
Finkelstein-Rubinstein constraints for a Skyrme configuration which is
only known numerically.  

\begin{enumerate}
\item Calculate the minimal energy configuration and analyze its symmetry 
properties.
\item Confirm the symmetry by starting with a symmetric configuration as 
initial condition and relaxing to the same final
configuration.\footnote{This provides a homotopy from the initial
  condition to the final configuration which is invariant under the symmetry.}
\item For each generator of the symmetry group identify $k$ for $C_n^k$.
\item Approximate the Skyrmion by a truncated rational map ansatz with
  the right symmetries. 
\item Calculate the Finkelstein-Rubinstein constraints using formula
  (\ref{generalN}). 
\end{enumerate}

\begin{figure}[!htb] 
\begin{center}
\includegraphics[height=50mm,angle=0]{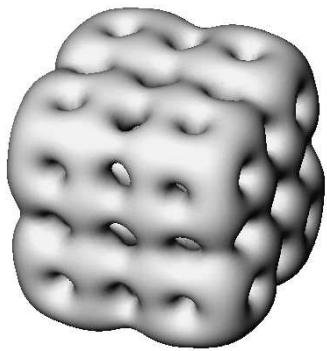}
\caption{The $B=32$ cube \label{Sk32}}
\end{center}
\end{figure}

In the following, we calculate the Finkelstein-Rubinstein constraints
for the $B=32$ cube, 
which is displayed in figure \ref{Sk32}. This configuration is one of the
first examples of a Skyrmion which cannot be described with the
rational map ansatz \cite{Battye:2004rw}. The configuration has been
calculated numerically in \cite{Battye:2004rw} (step 1). 
It can be approximated by a chunk of the Skyrmion
crystal \cite{Baskerville:1995db} and has cubic symmetry. Starting
with a chunk of the crystal as initial conditions imposes the symmetries
and corresponds to step 2. Step 3 is comparatively easy in this
example since we have an analytic ansatz for the initial condition.
The cubic symmetry is generated by a $C_3$ and a $C_4$ symmetry. 
The Skyrme field can be parametrized as $\sigma + i \tau_j \pi_j$
where $\tau_i$ are the Pauli matrices. With the choice of fields as in
\cite{Battye:2004rw} the symmetries are
\begin{equation}
\nonumber
\begin{array}{ll}
C_3: & (\sigma,\pi_1,\pi_2,\pi_3) 
\mapsto (\sigma, \pi_3,\pi_1, \pi_2), \\
C_4: & (\sigma,\pi_1,\pi_2,\pi_3)
\mapsto (\sigma,
\frac{-2 \pi_1 + \pi_2  - 2 \pi_3}{3},
\frac{\pi_1 - 2 \pi_2 - 2 \pi_3}{3},
\frac{-2 \pi_1 - 2 \pi_2 + \pi_3}{3}).
\end{array}
\end{equation}
So, the cubic symmetry is generated by $C_3^1$ and $C_4^2$. 
The $B=32$ Skyrmion can be thought of as eight $B=4$ cubes. Under the
$C_4^2$ symmetry the top four cubes transform into each other and the
bottom four cubes transform into each other (step 4). So, we can use formula
(\ref{generalN}) with $B_i = 4$ and $l=2$, $n=4$ and $k=2$ (step 5). 
There are no Skyrmions on the symmetry axis, ${\tilde B}_i = 0$.
\begin{eqnarray}
\nonumber
N &=& 2*4*(4*4-2), \\
\nonumber
  &=& 112.
\end{eqnarray}
Therefore, the Finkelstein-Rubinstein phase $\chi_{FR} = 1$ for this
symmetry.  
Under the $C_3^1$ symmetry, two $B=4$ Skyrmions are on the rotation
axes and there are two groups of three $B=4$ Skyrmions which transform
into each other. Therefore, $B_i = 4$, $l=2$, $n=3$, $k=1$, ${\tilde
  B} =4$ and $m=2$. So, formula (\ref{generalN}) gives
\begin{eqnarray}
\nonumber
N &=& 2*4*(4*3-1) + 2*4*(4-1)/3, \\
\nonumber
  &=& 96.
\end{eqnarray}
Therefore, the Finkelstein-Rubinstein phase is again trivial. In the
following section, we show that we can derive some results from
general principles, so that we only have to carry through steps 1 -- 5
for a very small subset of all the possible symmetries.

\section{Symmetries and Finkelstein-Rubinstein constraints for even $B$}
\label{evenB}

In this section we collect a set of general results about symmetries
and Finkelstein-Rubinstein constraints. We start with the following lemma.

\begin{lemma}
\label{Beven}
Negative Finkelstein-Rubinstein constraints cannot occur for
$C_{2l+1}^k$ for $l \ge 1$ if $B$ is even.
\end{lemma}

\noindent {\bf Proof:}
Applying a $C_n^k$ symmetry $n$ times corresponds to a $2 \pi$
rotation in space followed by a $2 \pi k$ rotation in target space. If
$B$ is even, then a $2 \pi$ rotation (and also a $2 \pi$ isorotation) 
is homotopic to the trivial loop in
the Skyrme configuration space. In this case,
the Finkelstein-Rubinstein constraints correspond to one-dimensional
and hence irreducible representations of $C_n^k$ which are obtained by
mapping the generator of $C_n^k$ to $(-1)^N$. This representation can
be thought of as a homomorphism from $C_n^k \to {\mathbb Z}_2$ and
therefore can only be nontrivial if $n$ is even. \hfill $\square$

In the following, we address the question of which symmetries can lead to
negative Finkelstein-Rubinstein phases. Lemma \ref{Beven} greatly simplifies
the discussion, so we only consider even baryon number
$B$. Furthermore, we restrict our attention to the symmetries which
have been found empirically \cite{Battye:2001qn}. These are cyclic
symmetry $C_2$, dihedral symmetry $D_n$ for $n \le 6$, the tetrahedral
group $T$, the octahedral group $O$ and the icosahedral group $Y$. 
Therefore, we first discuss the cyclic subgroups $C_n^k$ for $n \le 6$.
For even baryon number $B$ the following picture emerges for Skyrmions
which are well-approximated by rational maps. Using
formula (\ref{N}) and lemma \ref{Beven} the cyclic groups $C_n^k$ can be 
grouped into three groups: 

\begin{enumerate}
\item The following symmetries never lead to negative $\chi_{FR}$, 
if $B \equiv 0 \mod 2$: \\
$C_2^0,~C_3^k,~C_4^2,~ C_5^k,~ C_6^{2k}$.

\item The following symmetries give rise to negative $\chi_{FR}$ 
if $B \equiv 2 \mod  4$: \\
$C_2^1,~C_4^0,~C_6^{2k+1}$.

\item The following symmetries give rise to negative $\chi_{FR}$ if $B
  \equiv 4 \mod 8$: \\
$C_4^1,~C_4^3$.
\end{enumerate}

In order to understand $D_n$ symmetry we need to examine 
under which conditions can there be an additional $C_2$ symmetry for a
given realization of a $C_n^k$ symmetry? 
Since $2 \pi$ isorotations are always a symmetry we can restrict our
attention to $j=0$ and $j=1$. 
The two different realizations of a $C_n^k$ symmetry can be
characterized by their zeros and poles at zero and infinity. 
A rational map of type (\ref{rat1}) has a zero of multiplicity 
$k \mod n$ at $z=0$ and a pole of multiplicity $k \mod n$ at
$z=\infty$.
A rational map of type (\ref{rat2}) has a pole of multiplicity $n-k
\mod n$  at $z=0$, and a pole of order $k \mod n$ at $z=\infty$.

For our purpose, it is sufficient to discuss the case that the $C_2$
rotation axis is orthogonal to the $C_n^k$ rotation axis. 
This generates the group $D_n$. 
For a $C_2^1$ symmetry it is important
whether the $C_2^1$ isorotation axis is parallel to the $C_n^k$
isorotation or orthogonal to it, and we will introduce the notation 
$(C_2^1)_\parallel$ and $(C_2^1)_\perp$. For $C_2^0$ such a
distinction is not necessary. 
A $C_2$ symmetry around an axis
orthogonal to the 
$C_n^k$ symmetry axis maps $z=0$ to $z=\infty$.  
In target space, $R=0$ is mapped to $R=\infty$ if the $C_2^1$ axis is
orthogonal to the $C_n^k$ axis in target space, namely for $(C_2^1)_\perp$, 
but $R=0$ and $R=\infty$ are invariant for $(C_2^1)_\parallel$ and for
$C_2^0$. 
The numbers of zeros and poles imply
that a rational map of type (\ref{rat1}) cannot have an additional
symmetry of type $C_2^0$ or $(C_2^1)_\parallel$. However, a
$(C_2^1)_\perp$ symmetry is possible for all values of $k$. 
A rational map of type (\ref{rat2}) 
can only have an additional $C_2^0$ or $(C_2^1)_\parallel$ 
symmetry if $n-k \equiv k \mod n$. 
An additional $(C_2^1)_\perp$ symmetry is never allowed.  

Now, we can apply the above result to the case $B \equiv 0 \mod
4$. The above list shows that a negative Finkelstein-Rubinstein phase
is only possible for $C_4^1$ and $C_4^3$ symmetry. 
A $C_4^k$ symmetry empirically only occurs as a subgroup of a
$D_4$ symmetry which itself might be a subgroup of the cubic group
$O$.  
For $B \equiv 0 \mod 4$ the corresponding rational map
can be of type (\ref{rat1}). Then an additional $C_2$ symmetry is only
possible for $4 - k \equiv k \mod 4$ which excludes $k=1$ and
$k=3$. The rational map can also be of type (\ref{rat2}) provided that
$k \equiv 0 \mod 4$. 
Therefore, $D_4$ symmetry is not compatible with $C_4^1$ and
$C_4^3$ so that no negative Finkelstein-Rubinstein constraints can
occur for $B \equiv 0 \mod 4$. 

Now, we discuss the case $B \equiv 2 \mod 4$. Negative $\chi_{FR}$ can
occur for $C_2^1$, $C_4^0$ and $C_6^{2k+1}$. $C_2^1$ can occur as
subgroup of $D_n$, $T$, $O$ and $Y$. From the point of view of
Finkelstein-Rubinstein constraints only the $D_2$ subgroup of $T$ and
$Y$ contributes and similarly, only the $D_4$ subgroup of $O$
contributes. 
Rational maps of type (\ref{rat1}) always allow a $(C_2^1)_\perp$ symmetry
which leads to $\chi_{FR} = -1$. For rational maps of type
(\ref{rat2}), the $C_2$ symmetry is either $C_2^0$ or
$(C_2^1)_\parallel$, and only the latter leads to negative
$\chi_{FR}$. Note however, that a $D_2$ symmetry of type (\ref{rat2})
is only possible if $k \equiv 1 \mod 2$, so that a $D_2$ symmetry
always leads to negative $\chi_{FR}$. 
In \cite{Krusch:2002by}, the symmetries and Finkelstein-Rubinstein
phases have been discussed for $B \le 22$. The results for 
$B \equiv 2 \mod 4$ are displayed in table \ref{table1}. 
Note that type (\ref{rat2}) can only occur if $B
\equiv 0 \mod n$ for the maximal $C_n$ subgroup. 
All symmetries are  either $D_2$ or are of type (\ref{rat1}),
so that for $B \equiv 2 \mod 4$ only symmetries with negative
Finkelstein-Rubinstein phases have been observed. However, from
symmetry arguments alone, it is not possible to exclude that the
symmetries act in such a way that all the phases are positive.

\begin{table}
\begin{center}
\begin{tabular}{|c|c|c|c|}
\hline
&&&\\
$B$ & symmetry & type & $\chi_{FR}$\\
&&&\\
\hline
&&& \\
$6$     & $D_4$ & (\ref{rat1}) & $(-1)$ \\
$10$    & $D_4$ & (\ref{rat1}) & $(-1)$\\
$10^*$  & $D_3$ & (\ref{rat1}) & $(-1)$\\
$14$    & $D_2$ & (\ref{rat1}) or (\ref{rat2}) & $(-1)$\\
$18$    & $D_2$ & (\ref{rat1}) or (\ref{rat2}) & $(-1)$\\
$22$    & $D_5$ & (\ref{rat1}) & $(-1)$\\
$22^*$  & $D_3$ & (\ref{rat1}) & $(-1)$\\
&&&\\
\hline
\end{tabular}
\caption{Symmetry, type and Finkelstein-Rubinstein phase for baryon
  number $B \equiv 2 \mod 4$.\label{table1}}
\end{center}
\end{table}

\subsection{Physical interpretation of the symmetry calculations}

The physical interpretation of this result is as follows. In nuclear
physics we are interested in the quantum ground states for a given
number of nucleons. In our semi-classical approximations the ground
state is given by the lowest values of the angular momentum quantum
numbers $J$ and $I$ which are compatible with the symmetries of the
classical configurations. In particular, the wave function $\psi$ has
to satisfy the symmetry condition (\ref{condition}) for all classical
symmetries of the given Skyrmion. We decompose the wave function
$\psi$ in angular momentum eigenfunction and write $\psi = |J\rangle
|I \rangle$ for a wave function with angular momentum quantum numbers
$J$ and $I$. 

Particularly interesting is the even-even situation when there is an
  even number of protons and an even number of neutrons. For small
  nuclei the number of protons and neutrons are equal, so that
  even-even nuclei have  $B\equiv 0 \mod 4$. 
In this case, experiment shows that the ground
  state is generally given by $|0\rangle |0\rangle$. This is only
  possible if the Finkelstein-Rubinstein constraints are
  trivial. The 
above discussion showed that the $|0\rangle|0\rangle$ ground state
is allowed for $B \equiv 0 \mod 4$, in agreement with experiment.
In this calculation we assumed that the relevant Skyrmions are
well-approximated by rational maps, and that $C_4$ symmetry only
occurs as a subgroup of a $D_4$ symmetry (which might itself be a
subgroup of an octahedral symmetry). 

For higher pion mass the
configurations deviate significantly from the rational map ansatz
\cite{PC1},
which raises the question whether it is possible to classify
symmetries allowing for the more general truncated rational map
ansatz. The Skyrmions would be allowed to split into groups and the
Finkelstein-Rubinstein constraints have to be calculated with formula
(\ref{generalN}). Note that if Skyrmions can be thought of as being
composed only of $B=4$ Skyrmions ($\alpha$ particles) then formula
(\ref{generalN}) and the above symmetry discussion also imply that
there are no negative Finkelstein-Rubinstein phases, so that the
ground state for such Skyrmions is $|0\rangle |0\rangle$.

For odd-odd nuclei, which implies $B \equiv 2 \mod 4$ for equal number
of protons and neutrons, experiment shows that the ground state is
usually not given by $|0\rangle|0\rangle$. This is consistent with our
observation that negative Finkelstein-Rubinstein phases occur.
However, symmetry arguments alone are not sufficient to prove the
occurrence of negative Finkelstein-Rubinstein phases. For odd baryon
number, lemma \ref{Beven} cannot be applied and there is little hope
of finding simple rules.

\section{Conclusion}

In this paper we discussed how to calculate Finkelstein-Rubinstein
constraints for configurations which are only known numerically. 
This is an important problem since recent calculations show that
for large pion mass Skyrmions are not very well described by rational
maps \cite{PC1}. Moreover, there is mounting evidence that the 
pion mass in the Skyrme model should be interpreted as 
an effective mass with a value at least twice the physical value 
\cite{Battye:2005nx}.
The calculation of the Finkelstein-Rubinstein constraints can be
incorporated into the algorithm for finding the minimal energy
configurations with only minor modifications. 

Note that for a wide range of values of the pion mass, the
rational map 
ansatz works well for small enough baryon number $B$. Therefore, the
truncated rational map ansatz in its current form has a good chance of
capturing all the physically relevant Skyrmions. An obvious but
slightly tedious generalisation would be to allow for the constituent
Skyrmions to be themselves approximated by a truncated rational map
ansatz. 

The paper also discussed which symmetries occur for rational map
Skyrmions and when to expect negative Finkelstein-Rubinstein
constraints for even baryon numbers. In particular, the Skyrme model
calculations suggest the correct phenomenological trend, namely that
the ground state of even-even nuclei have the ground state 
$|0 \rangle |0 \rangle$.

\section*{Acknowledgements}

The author would like to thank N. S. Manton, J. M. Speight and
P. M. Sutcliffe for many fruitful discussions. The author also 
acknowledges an EPSRC Research fellowship GR/S29478/01.

\label{lastref}


\begin{thebibliography}{10}

\bibitem{Skyrme:1961vq}
T.~H.~R. Skyrme, {\em A nonlinear field theory\/}, Proc. Roy. Soc. Lond. {\bf
  A260}: 127 ({\bf 1961}),

\bibitem{Battye:2001qn}
R.~A. Battye and P.~M. Sutcliffe, {\em Skyrmions, Fullerenes and Rational
  Maps\/}, Rev. Math. Phys. {\bf 14}: 29 ({\bf 2002}),
  {\ttfamily{<hep-th/0103026>}},

\bibitem{Houghton:1998kg}
C.~J. Houghton, N.~S. Manton and P.~M. Sutcliffe, {\em Rational Maps, Monopoles
  and Skyrmions\/}, Nucl. Phys. {\bf B510}: 507 ({\bf 1998}),
  {\ttfamily{<hep-th/9705151>}},

\bibitem{Battye:2004rw}
R.~Battye and P.~Sutcliffe, {\em Skyrmions and the pion mass\/}, Nucl. Phys.
  {\bf B705}: 384--400 ({\bf 2005}), {\ttfamily{<hep-ph/0410157>}},

\bibitem{Adkins:1983ya}
G.~S. Adkins, C.~R. Nappi and E.~Witten, {\em Static properties of nucleons in
  the {S}kyrme model\/}, Nucl. Phys. {\bf B228}: 552 ({\bf 1983}),

\bibitem{Adkins:1984hy}
G.~S. Adkins and C.~R. Nappi, {\em The {S}kyrme model with pion masses\/},
  Nucl. Phys. {\bf B233}: 109 ({\bf 1984}),

\bibitem{Finkelstein:1968hy}
D.~Finkelstein and J.~Rubinstein, {\em Connection between spin, statistics, and
  kinks\/}, J. Math. Phys. {\bf 9}: 1762 ({\bf 1968}),

\bibitem{Braaten:1988cc}
E.~Braaten and L.~Carson, {\em The Deuteron as a toroidal Skyrmion\/}, Phys.
  Rev. {\bf D38}: 3525 ({\bf 1988}),

\bibitem{Verbaarschot:1987au}
J.~J.~M. Verbaarschot, {\em Axial symmetry of bound baryon number two solution
  of the {S}kyrme model\/}, Phys. Lett. {\bf B195}: 235 ({\bf 1987}),

\bibitem{Leese:1995hb}
R.~A. Leese, N.~S. Manton and B.~J. Schroers, {\em Attractive channel Skyrmions
  and the Deuteron\/}, Nucl. Phys. {\bf B442}: 228 ({\bf 1995}),
  {\ttfamily{<hep-ph/9502405>}},

\bibitem{Carson:1991yv}
L.~Carson, {\em B = 3 nuclei as quantized Multiskyrmions\/}, Phys. Rev. Lett.
  {\bf 66}: 1406 ({\bf 1991}),

\bibitem{Walhout:1992gr}
T.~S. Walhout, {\em Quantizing the four Baryon Skyrmion\/}, Nucl. Phys. {\bf
  A547}: 423 ({\bf 1992}),

\bibitem{Irwin:1998bs}
P.~Irwin, {\em Zero mode quantization of Multi-skyrmions\/}, Phys. Rev. {\bf
  D61}: 114024 ({\bf 2000}), {\ttfamily{<hep-th/9804142>}},

\bibitem{Kopeliovich:2001yg}
V.~B. Kopeliovich, {\em Characteristic predictions of topological soliton
  models\/}, J. Exp. Theor. Phys. {\bf 93}: 435--448 ({\bf 2001}),
  {\ttfamily{<hep-ph/0103336>}},

\bibitem{Krusch:2002by}
S.~Krusch, {\em Homotopy of Rational Maps and the Quantization of Skyrmions\/},
  Ann. Phys. {\bf 304}: 103--127 ({\bf 2003}), {\ttfamily{<hep-th/0210310>}},

\bibitem{Faddeev:1976pg}
L.~D. Faddeev, {\em Some comments on the many dimensional solitons\/}, Lett.
  Math. Phys. {\bf 1}: 289 ({\bf 1976}),

\bibitem{Battye:2002wc}
R.~A. Battye, C.~J. Houghton and P.~M. Sutcliffe, {\em Icosahedral
  Skyrmions\/}, J. Math. Phys. {\bf 44}: 3543--3554 ({\bf 2003}),
  {\ttfamily{<hep-th/0210147>}},

\bibitem{Houghton:2001fe}
C.~J. Houghton and S.~Krusch, {\em Folding in the {S}kyrme model\/}, J. Math.
  Phys. {\bf 42}: 4079 ({\bf 2001}), {\ttfamily{<hep-th/0104222>}},

\bibitem{Krusch:2005bn}
S.~Krusch and J.~M. Speight, {\em Fermionic Quantization of Hopf Solitons\/}
  ({\bf 2005}), {\ttfamily{<hep-th/0503067>}},

\bibitem{Giulini:1993gd}
D.~Giulini, {\em On the possibility of spinorial quantization in the {S}kyrme
  model\/}, Mod. Phys. Lett. {\bf A8}: 1917 ({\bf 1993}),
  {\ttfamily{<hep-th/9301101>}},

\bibitem{PC1}
R.~A. Battye and P.~M. Sutcliffe ({\bf 2005}),
Private Communication.

\bibitem{Manton:2000kj}
N.~S. Manton and B.~M. A.~G. Piette, {\em Understanding Skyrmions using
  Rational Maps\/}  
 Proceedings of the European Congress of Mathematics (Barcelona, 2000)
 Vol. 1, Progress in Mathematics 201, 469-479, Birkh\"auser, Basel,
({\bf 2001}), {\ttfamily{<hep-th/0008110>}},

\bibitem{Baskerville:1995db}
W.~K. Baskerville, {\em Making nuclei out of the {S}kyrme crystal\/}, Nucl.
  Phys. {\bf A596}: 611--630 ({\bf 1996}), {\ttfamily{<nucl-th/9510047>}},

\bibitem{Battye:2005nx}
R.~A. Battye, S.~Krusch and P.~M. Sutcliffe, {\em Spinning {S}kyrmions and the
  {S}kyrme parameters\/}  ({\bf 2005}), {\ttfamily{<hep-th/0507279>}}.

\end{thebibliography}
\end{document}